\documentclass[11pt,letterpaper]{article}
\usepackage{amsmath,amsfonts,latexsym}
\usepackage[colorlinks=false]{hyperref} 
\usepackage{fullpage}

\newcommand{\ep}{\epsilon}
\newcommand{\ket}[1]{\left| #1\right\rangle}        
\newcommand{\bra}[1]{\left\langle #1\right|}        
\newcommand{\kets}[1]{|\,#1\rangle}        
\newcommand{\bras}[1]{\langle #1|}        
\newcommand{\half}{\frac{1}{2}}
\newcommand{\ii}{\mathbb{I}}
\newcommand{\E}{{\cal E}}
\newcommand{\normi}[1]{\left\| #1\right\|_\infty}        
\newcommand{\normp}[1]{\left\| #1\right\|_p} 
\newcommand{\COMMENT}[1]{}
\newcommand{\normtwo}[1]{\left\| #1\right\|_2} 

\newtheorem{theorem}{Theorem}

\newtheorem{definition}{Definition}

\newenvironment{proof}
{\noindent {\bf Proof. }}
{{\hfill $\Box$}\\
 \smallskip}

\begin{document}


\title{On the Optimality of Quantum Encryption Schemes}
\author{Iordanis Kerenidis \\ Dept. of Mathematics, MIT \\jkeren@math.mit.edu
\and
Daniel Nagaj\\ Center for Theoretical Physics, MIT\\nagaj@mit.edu 
}


\maketitle

\begin{abstract}
It is well known that $n$ bits of entropy are necessary and sufficient to perfectly encrypt $n$ bits (one-time pad). Even if we allow the encryption to be approximate, the amount of entropy needed doesn't asymptotically change. However, this is not the case when we are encrypting quantum bits.
For the perfect encryption of $n$ quantum bits, $2n$ bits of entropy are necessary and sufficient (quantum one-time pad), but for approximate encryption one asymptotically needs only $n$ bits of entropy. 
In this paper, we provide the optimal trade-off between the approximation measure $\ep$ and the amount of classical entropy used in the encryption of single quantum bits. Then, we consider $n$-qubit encryption schemes which are a composition of independent single-qubit ones and provide the optimal schemes both in the 
$2$- and the $\infty$-norm. Moreover, we provide a counterexample to show 
that the encryption scheme of Ambainis-Smith \cite{AS04} based on
small-bias sets does not work in the $\infty$-norm.
\end{abstract}

\setcounter{page}{1}

\section{Introduction}

Secure transmission of information is a subject that has been studied
extensively. In this model, Alice wants to securely transmit a message
to Bob using a secret key that they both share, in such a way that any
eavesdropper gets absolutely no information about the message sent. In
the classical world, Shannon \cite{Sha48,Sha49} has shown that for the
perfect encryption of $n$ classical bits, it is necessary and
sufficient to use $n$ bits of classical entropy (one-time pad). By 
performing a bitwise XOR between the $n$-bit message and the $n$-bit 
secret key, the view of any eavesdropper that has no knowledge of the 
key is just a uniformly random $n$-bit string. 
Ambainis, Mosca, Tapp and de\,Wolf \cite{AMTW00} 
showed that $2n$ classical bits of entropy are necessary and
sufficient for the transmission of $n$ quantum bits. 

Let us briefly sketch how one can perfectly encrypt a quantum bit. Let
$\rho$ be the state of an arbitrary qubit and let $I,X,Y,Z$ be the
four Pauli matrices. Then, by using two bits of classical entropy we
can uniformly pick one of the four matrices and apply it to our
qubit. The state of the qubit after the encryption is
\[ \E(\rho) = \frac{1}{4} (\rho + X \rho X + Y \rho Y + Z \rho Z)\]
It's easy to verify that for all states $\rho$, $\E(\rho)= \half \ii$
and hence the view of the eavesdropper is the completely mixed state,
i.e. she gets no information about the encrypted state $\rho$. The
scheme easily generalizes to $n$-qubit states by using $2n$ classical
bits of entropy. 

The entropy needed for the perfect encryption of quantum states is
two times what is needed for the perfect encryption of classical
bits. Interestingly, this is no longer true, when we look at
{\em approximate encryption}.   
Let $\rho \in {\mathbb C}^{d \times d}$ be the state of a $(\log d)$-qubit message, $\{U_k \in {\mathbb C}^{d \times d}|k\in [N]\}$ be a set of $N$ unitary
operations acting on $\log d$ qubits and
${\cal D} = \{w_1,\ldots, w_N\}$ be a distribution on $[N]$. Imagine the encryption scheme, where Alice picks a unitary $U_k$ with probability $w_k$ and applies it to the message. The ciphertext can be written as  
\[ \E(\rho) =  \sum_{k\in [N]} w_k U_k \rho U_k^\dagger   \] 
and the entropy of the scheme is defined as the Shannon entropy $H({\cal D})$.  
\begin{definition}
The map $\E$ is a $(\epsilon, H)$-approximate
encryption scheme for the $\infty$-norm, if the entropy of the scheme is $H$
and for all states $\rho$  
\[  \normi{\E(\rho)- \frac{\ii}{d}} \leq \frac{\epsilon}{d}. \]
Similarly, the map $\E$ is a $(\epsilon, H)$-approximate
encryption scheme for the 2-norm, if the entropy of the scheme is $H$
and for all states $\rho$  
\[  \normtwo{\E(\rho)- \frac{\ii}{d}} \leq \frac{\epsilon}{\sqrt{d}}. \]
\end{definition}

Hayden, Leung, Shor and Winter \cite{HLSW04}
found an $(\ep, n+o(n))$-approximate encryption scheme for $n$ qubits. 
Specifically, they showed that an
encryption scheme that applies a unitary on $\rho$ picked uniformly
from a random set of unitaries of size $2^{n+o(n)}$ achieves
$\ep$-approximation.   
Ambainis and Smith \cite{AS04} derandomized this
construction using small-bias sets and constructed deterministically
a set of $2^{n+o(n)}$ unitaries that achieves an
$(\ep, n+o(n))$-approximation for the 2-norm.

On the other hand, it is not hard to see that for the classical case,
one needs at least $n-\log(1-\frac{\ep}{2})$ bits of entropy for an
$\ep$-approximation scheme. Hence, the entropy needed for the
approximate encryption of classical and quantum states is
asymptotically equal.    

In this paper, we start by investigating the approximate encryption of
single qubits and find the optimal trade-off between the approximation
measure $\ep$ and the amount of classical entropy $H$,
i.e. we calculate the least amount of classical entropy which is 
necessary and sufficient to achieve an $\ep$-approximation. Our proof
is constructive in the sense that for any given $\ep$ we 
describe the encryption scheme that achieves the optimal $H$
and vice versa. The following theorem holds both for the $\infty$- and
2-norm. Note the weights in the distributions are in decreasing order. 

\begin{theorem}
Let $\E(\rho)$ be the optimal $(\ep,H)$-approximate
encryption scheme for a qubit. Then,
\begin{enumerate}
\item The encryption is of the form $\E(\rho)= w \rho + x X \rho X
+ y Y \rho Y + z Z \rho Z$. 
\item For any fixed $\ep$, the optimal distribution $\cal D$ (and
hence the minimum entropy $H$) is:  
\begin{eqnarray*}
\begin{array}{rll}
   (i) & \ep \leq 1/6 & :\, {\cal D}=\{\frac{1}{4}+\frac{\ep}{2},\frac{1}{4}+\frac{\ep}{2},
													 \frac{1}{4}+\frac{\ep}{2}, \frac{1}{4}-\frac{3\ep}{2}\}, \\
  (ii) & 1/6 \leq \ep \leq 0.287 & :\, {\cal D}= \{2\ep,\half-\ep,\half -\ep,0\}, \\ 
 (iii) & \ep \geq 0.287 & : \, {\cal D} = \{\frac{1}{4}+\frac{3\ep}{2}, \frac{1}{4}-
 								 \frac{\ep}{2},\frac{1}{4}-\frac{\ep}{2},\frac{1}{4}- \frac{\ep}{2} \}.
\end{array}
\end{eqnarray*}
\end{enumerate}
\end{theorem}
In Section \ref{sectionPauli} we find the optimal Pauli encryption scheme for 
a qubit and in Section \ref{sectionOptimal} we show that Pauli encryption schemes 
are no worse than general encryption schemes. 

Next, in Section \ref{sectionNqubit} we consider $n$-qubit encryption
schemes which are a composition of independent single-qubit schemes that each 
use entropy $H$. In general, such questions are not easy to tackle, since they 
hinge on notoriously hard questions on the additivity of quantum channels. 
However, in this case we only consider unitary operations and hence we can use 
a result of King \cite{King} in order to find the optimal schemes. 
\begin{theorem}
Let $P$ be the single-qubit Pauli encryption scheme, which achieves
the optimal approximation $\ep$ for the given entropy $H$. Then, the optimal 
$n$-qubit independent encryption scheme $R(\rho)$ is the same for both the 
2- and the $\infty$-norm and has the following properties:
\begin{enumerate}
\item $R(\rho) = P^{\otimes n} (\rho)$.  
\item $\normtwo{R(\rho)-\frac{\ii}{2^n}}
\leq \sqrt{n}\frac{\epsilon}{2^{n/2}} +\frac{o(\epsilon\sqrt{n})}{2^{n/2}}.$
\item $\normi{R(\rho)-\frac{\ii}{2^n}}
\leq n\frac{\epsilon}{2^n} + \frac{o(n\epsilon)}{2^n}.$
\end{enumerate}
\end{theorem}

\noindent
The above bounds are tight and hence for any encryption scheme that acts
independently on each qubit, $2n-o(n)$ bits of entropy are
necessary for approximate encryption. 

Finally, in Section \ref{sectionNgeneral} we discuss non-independent
$n$-qubit encryption schemes. In particular,   
we are interested in the Ambainis-Smith small-bias set based scheme. In 
\cite{AS04}, they found a $(\epsilon, n+o(n))$-approximate encryption scheme 
for the 2-norm. Their scheme uses a deterministically constructed 
small-bias set of $2n$-bit strings of size $2^{n+o(n)}$, where each
string corresponds to a unitary which is a tensor product of $n$ Pauli
matrices. The message is encrypted by picking uniformly a unitary from
this set. One of the open questions in their paper is whether this
scheme is also an $(\epsilon, n+o(n))$-approximate
encryption scheme for the $\infty$-norm. We resolve this 
by finding an example of an asymptotically optimal small-bias set, 
for which the encryption scheme of Ambainis-Smith fails in the $\infty$-norm. 
However, it is possible that an $(\epsilon, n+o(n))$-approximate
encryption scheme for the $\infty$-norm can be constructed in a different way, 
for example by using a small-bias set with some extra properties.


\section{The Optimal Pauli Encryption Scheme} 
\label{sectionPauli}

The input state to our encryption scheme is a quantum bit which can be
described by a density matrix $\rho$, i.e. a hermitian matrix with
unit trace
\begin{eqnarray}
	\rho = \frac{1}{2}\left(\ii + r_x X + r_y Y + r_z Z\right), \label{densitymatrix}
\end{eqnarray}
where $\vec{r}=(r_x,r_y,r_z)$ is a unit vector, and 
the four Pauli matrices are
\[ 
\ii = 
\left[\begin{array}{cc}
	1 & 0 \\ 
	0 & 1
\end{array}\right], \quad
X = 
\left[\begin{array}{cc}
	0 & 1 \\ 
	1 & 0
\end{array}\right], \quad
Y = 
\left[\begin{array}{cc}
	0 & -i \\ 
	i & 0
\end{array}\right], \quad
Z = 
\left[\begin{array}{cc}
	1 & 0 \\ 
	0 & -1
\end{array}\right].
\]
Let us denote $+1$ eigenvectors of the matrices $X,Y$ and $Z$
by $\ket{x+},\ket{y+}$ and $\ket{z+}$.

A {\em Pauli encryption scheme} for single qubits is described by a
probability distribution on the four Pauli matrices, i.e. by a
probability vector ${\cal D}=(w,x,y,z)$, such that the encryption of a
qubit $\rho$ is given by 
\begin{equation}
\label{encoding}
\E_{IXYZ}(\rho) =  w \rho + x X \rho X  + y Y\rho Y + z Z \rho Z.
\end{equation}
Without loss of generality, we can assume the weights $\{w,x,y,z\}$
obey $w\geq z \geq x \geq y \geq 0$. The reason for this is that these four
unitaries are freely interchangeable by picking a suitable
$\rho=U\rho'U^{\dagger}$. If the original qubit $\rho$ was encoded by
$\E(\rho)$, with weights $\{w,x,y,z\}$, we can achieve the same
encoding $E'(\rho')$ on the transformed qubit $\rho'$, just with
$\{w,x,y,z\}$ permuted.   

The classical entropy used by the encryption scheme is the entropy of
the probability distribution, i.e. $H(\{p_i\})=-\sum_i p_i \log_2 p_i =  -w\log w
-x\log x- y\log y -z\log z$.

To test how good the encryption scheme is, we want to know how much the
encrypted state differs from the completely mixed state in the 2- and the operator 
norm. For any $d$-dimensional matrix $A$, the 2- and operator norm are related 
to the eigenvalues of the matrix, namely
\[
	\normtwo{A}^2
		\;=\; \sum_{k=1}^{d} \lambda_k^2, \;\;\;
	\normi{A} 
		\;=\; \max_k |\lambda_k|	
\]
Thus, for the operator norm we need to examine the maximum of the absolute value
of the eigenvalues of
\[
I(\rho)=\E_{IXYZ}(\rho) - \half\mathbb{I}.
\]
Note, that since the matrix $I(\rho)$ has trace equal to 0, the two eigenvalues 
are of the form $\pm \lambda$ and hence, the 2-norm is maximized simultaneously 
with the operator norm.


\subsection{The maximum eigenvalue of $I(\rho)$}
\label{subsecMaxeigenvalue}

After applying the channel (\ref{encoding}) to the density matrix
described by (\ref{densitymatrix}), we obtain
\[	
	\E_{IXYZ}(\rho) =  \frac{1}{2}\left(\ii + r'_x X + r'_y Y + r'_z Z\right),
\]
where the new parameters can be easily determined from (\ref{encoding}) using 
the anticommutation relations for Pauli matrices.
\begin{eqnarray*}
	r'_x &=& (w+x-z-y)r_x = (2(w+x)-1) r_x, \\
	r'_y &=& (w+y-z-x)r_y = (2(w+y)-1) r_y, \\
	r'_z &=& (w+z-x-y)r_z = (2(w+z)-1) r_z.
\end{eqnarray*}
This shows that the parameters $r_x, r_y$ and $r_z$ shrink according
to the above relations. The factors can be negative, but because
have $w\geq z\geq x \geq y$ and $w+z+x+y=1$, with a little work 
one can verify that the magnitude of the shrinking factor $|2(w+z)-1|$ in 
front of $r_z$ is the largest of the three.  

Using the geometric description (\ref{densitymatrix}) of $\rho$, we can express 
the matrix $I(\rho)$ as 
\[
	I(\rho)=\E(\rho)-\tfrac{1}{2}\ii = \frac{1}{2} \left( r'_x X + r'_y Y + r'_z Z \right).
\]
Its eigenvalues are then simply 
\[
	\lambda_{I(\rho)}=\pm\frac{1}{2}|\vec{r}'|.
\]
Our goal is to find the maximum eigenvalue $|\lambda_{I(\rho)}|$ 
over all states $\rho$ as a function of the
probability distribution ${\cal D}=(w,z,x,y)$ and then pick the
distribution that minimizes it. Already knowing that the shrinking factor in front of 
$r_z$ is the largest, we can maximize $|\lambda_{I(\rho)}|=\tfrac{1}{2}(1+|\vec{r}'|)$
by picking $\rho$ with $\vec{r}=(0,0,1)$. This gives us $\vec{r}'=(0,0,2(w+z)-1)$,
and 
\begin{eqnarray}
	max_\rho|\lambda_{I(\rho)}| = \left|w+z-\half\right|, \label{precision}
\end{eqnarray}
Note that $w$ and $z$ are the two largest weights and therefore we always have $w+z\geq\frac{1}{2}$.

We thank the referees for pointing out a geometric view of the Pauli encoding in \cite{NCbook}, 
which simplified the proof in this section.


\subsection{The optimal trade-off between approximation and entropy}

In Section \ref{subsecMaxeigenvalue}, we found an upper bound on the maximum
eigenvalue of $I(\rho)$ as a function of the probability distribution
used by the Pauli encryption scheme. Note also that equation
(\ref{precision}) shows that for a perfect encryption the only
possible scheme is the one that uses a uniform distribution over the
four Pauli matrices. 
 
The natural question is to find the optimal Pauli encryption scheme
when we can only use a fixed amount $H$ of classical entropy. Turning
the question around, we fix the approximation parameter $\ep$ and
calculate the necessary entropy to achieve it. 

Let us fix 
$\ep=max_\rho|\lambda|=w+z-1/2$.
In addition, the condition $w\geq z$ implies that $1/4+\ep/2 \geq z$.
Our goal is to minimize the classical entropy needed to achieve
approximation $\ep$:
\begin{eqnarray*}
H(\ep)=min_{{\cal D}} \left(-w \log w  -z \log z -x \log x -y \log y \right).
\end{eqnarray*}
Keeping $x$, $y$ and $\ep$ fixed, the entropy as a function of $z$ is
concave, with a maximum at $z=w$. Because $z\leq w$, the entropy
decreases with decreasing $z$. Specifically, if $z>x+y$, one can decrease 
the entropy by setting $z=x+y$ (and increasing $w$ accordingly, to keep $\ep$ fixed). 
Without loss of generality, one can then assume that $z \leq x+y$ 
for the optimal ${\cal D}$. 
Now, let us minimize the entropy as a function of $x$. 
It is concave in $x$, with a maximum at $x=y=(1-w-z)/2$ 
and possible minima at the endpoints. Because $x\geq y$, we want to pick $x$ 
as large as possible. Because $x\leq z$ and $z\leq x+y$, 
this results in $x=z$.
The weights that minimize the entropy for a fixed $\ep$ thus are (as a function of $z$)
\[
  w\; = \;1/2+\ep-z, \quad x\;=\;z, \quad y\; = \; 1/2-\ep-z. 
\]
To find the optimal $H(\ep)$, one thus needs to minimize 
\begin{eqnarray*}
H(\ep,z)=-\left(\half+\ep-z\right)\log\left(\half+\ep-z\right)
-2z \log z -\left(\half-\ep-z\right)\log\left(\half-\ep-z\right).\nonumber
\end{eqnarray*}
with respect to $z$, remembering the constraints collected 
so far ($w\geq z \geq x \geq
y \geq 0$, $x+y \geq z$):
\begin{eqnarray}
\frac{1}{4}+\frac{\ep}{2} &\geq& z \geq \frac{1}{4}-\frac{\ep}{2},
\label{zconstraint1}\\ 
\half-\ep &\geq& z. \label{zconstraint2}
\end{eqnarray}
We perform this minimization in Appendix A.1 and conclude that
for $\ep\leq 1/6$, picking the three larger
weights to be equal is the entropy-minimizing strategy. 
For $1/6 \leq \ep \leq \ep_0$,
picking only three unitaries, with two of the lower weights equal is
the best choice. For $\ep_0 \leq \ep \leq 1/2$, it is optimal to 
pick the three smaller weights to be equal.

Turning the argument around -- given entropy $H$, what is the
optimal Pauli encryption scheme? There is a unique way to pick the
probability distribution with the given entropy that minimizes the
parameter $\ep$. 
\begin{eqnarray}
\begin{array}{rll}
 (i) & H \geq {\log_2}3 & : \,{\cal D}=(z,z,z,1-3z), \\
 (ii) & {\log_2}3 \geq H \geq H_0 & : \,{\cal D}=(1-2z,z,z,0), \\
 (iii) & H_0 \geq H & : \,{\cal D}=(1-3z,z,z,z). 
\end{array}
\label{distrib}
\end{eqnarray}
Note that the weights are in descending order and that the
approximation $\ep$ is given by the sum of the largest two weights
minus $\half$. 
Also, one should not expect the optimal distribution parameters 
to be continuous at $H_0$. These two ways of picking the weights 
come from different regions in the parameter space $\{w,z,x,y\}$, 
and the choice of the optimal distribution is simply a numerical 
minimum of these two functions. The point $H_0$ (or equivalently  
$\ep_0$) does not have an obvious special meaning.


\section{The optimality of Pauli Encryption Schemes} 
\label{sectionOptimal}

In this section we give an elementary constructive proof 
that the Pauli encryption schemes are no worse
than any general encryption scheme. For
any encryption scheme $\E(\rho) = \sum_k p_k U_k \rho U^\dagger_k$ with
arbitrary unitaries and weights, we give a Pauli encryption scheme
with weights $\{w,z,x,y\}$ that has lower entropy, and is no worse
than $\E(\rho)$. We show this by finding a density matrix $\rho_0$,  
for which the maximum eigenvalue of $I(\rho_0)$ is the same as in
(\ref{precision}), which is the worst case for the newly found Pauli scheme.
Hence, the Pauli encryption scheme of section
\ref{sectionPauli} is optimal amongst all possible encryption schemes.

After completion of this work, we learned of an alternative proof of
optimality of Pauli encryption for a single qubit by Bouda and Ziman  
\cite{BoudaZiman}. They investigated perfect encryption of a subspace
of the Bloch sphere, while we are interested in approximate encryption  
of the whole Bloch sphere. Their proof uses the Kraus representation
of the quantum channel, showing that the representation of a channel
using orthogonal matrices requires the least amount of entropy. We 
also thank the anonymous referee for providing us with another shorter proof. Using
the fact that every channel can be expressed also as a Pauli channel (\cite{Ruskai}),
we can utilize a clever trick by Nielsen (\cite{Nielsen1}) 
to prove that the weights of this Pauli
channel majorize the weights of the original channel. Knowing that
the entropy is concave, we can conclude that the Pauli realization
of the channel requires the least amount of entropy. The details of this 
proof are given in Appendix A.2. Let us now continue with our proof.

Let $T$ be an encryption scheme with distribution
$\{w_1,w_2,\ldots,w_N\}$ over $N$ unitaries $U_k$, where the weights are
in decreasing order.  
We parametrize the unitaries as $U_k =e^{i\alpha_k}e^{i\phi_k
(\vec{n}_k \cdot \vec{\sigma})}$, where $\vec{n}_k=(x_k, y_k, z_k)$ and $\vec{\sigma}=(X,Y,Z)$. 
The phases $\alpha_k$ are not important in our analysis and hence, we denote the
parametrization of $U_k$ only as $U(\phi_k,\vec{n}_k)$. 

We have the following three cases:\\

\noindent
{\bf Case 1:} $w_1+w_2-1/2 \leq 0$

We show that the entropy $H$ of the encryption scheme $T$ 
is greater or equal to 2, and for $H=2$, we already know a perfect
encoding with four unitaries and $w_k=1/4$. 
It is clear, that if $w_1 < 1/4$ then the entropy is larger than
2. Let us assume that $w_1 \geq 1/4$. From the concavity of the
Shannon entropy, we know that the entropy of a distribution that
contains two weights $(w_k,w_l)$ with $w_k\geq w_l$ decreases if we
change them into $(w_k+\Delta,w_l-\Delta)$. 

Hence we can decrease the
entropy of the initial distribution $\{w_1,w_2,\ldots,w_N\}$ by
increasing the weight $w_2$ to make it equal to $w_2' = 1/2-w_1$ and
decreasing some of the smaller weights. We can further decrease the
entropy by making the middle weights all equal, i.e 
$\{w_1,w_2',w_2',\ldots,w_2',w_{N}\}$.
Picking $w_1 = x$ fully determines the distribution, giving 
$w_2' = 1/2-x$ and $w_N = (4-N)/2+(N-3)x$. The constraints
$1/2\geq w_1\geq w_2\geq w_N\geq 0$ give us: 
\begin{eqnarray}
\frac{N-3}{2(N-2)}\geq x \geq \frac{N-4}{2(N-3)}. \label{xconstraint}
\end{eqnarray}

The entropy as a function of $x$ is concave (the second derivative is
negative) and therefore, we look for the minimum entropy at the
endpoints, given in (\ref{xconstraint}). These endpoints correspond to
choosing the distribution as $\{w_1,w_2,\dots,w_2\}$ with
$w_2=(1-w_1)/(N-1)$. 
The entropy of such distributions as a function of $N$ is 
\begin{eqnarray*}
H(N)=-w_1 \log w_1 - (N-1) w_2 \log w_2 
= -\frac{N-3}{2(N-2)} \log (N-3) + 1 + \log (N-2).
\end{eqnarray*}
It's easy to see that this function is a monotone, growing function
of $N$ with a minimum for $H(4)=2$. 
We conclude that any encryption scheme with $n\geq 5$ unitaries
and $w_1+w_2-1/2<0$ uses entropy $H\geq 2$ and hence is worse than the
perfect encryption scheme with four unitaries.\\

\noindent
{\bf Case 2:} $w_1+w_2-1/2 \geq 0$ and $\sum_{k=3}^n w_k \leq 2w_2$. 

We show that there exists a Pauli encryption scheme that is no worse
than $T$ and uses less entropy. Let $P$ be the Pauli scheme that uses
the distribution $\{w_1,w_2,w_2,w'_3\}$. This is possible by the
constraint $\sum_{k=3}^n w_k \leq 2w_2$ and from the concavity of the
entropy, $P$ uses less entropy. We also know from equation (\ref{precision}) that for the encryption scheme $P$ 
\begin{eqnarray*}
\textrm{max}_\rho|\lambda_{I(\rho)}| = \left|w_1+w_2-\half\right|.
\end{eqnarray*}
Without loss of generality, when encoding an input density matrix $\rho$
with the set of unitaries
\begin{eqnarray*}
U(\phi_1,\vec{n}_1), U(\phi_2,\vec{n}_2), U(\phi_3,\vec{n}_3),\ldots,
U(\phi_n,\vec{n}_n) \, 
\end{eqnarray*}
one can equivalently analyze the encoding of the density matrix 
$\rho'=U_1^{\dagger} \rho U_1$ with a related set of unitaries:
\begin{eqnarray*}
{\ii}, U(\phi_2',\vec{n}_2'),
U(\phi_3',\vec{n}_3'),\ldots, U(\phi_n',\vec{n}_n'). 
\end{eqnarray*}
The approximation parameter $\ep$ of the encoding scheme is basis independent, 
it is now convenient to pick a basis in which the unitaries are of the form  
\begin{eqnarray*}
 {\mathbb{I}}, Z_{\alpha_2}, \left(z_3 Z+ x_3 X+y_3
Y\right)_{\alpha_3},\ldots, 
\left(z_n Z+x_n X+y_n Y\right)_{\alpha_n}, \label{pickUs} 
\end{eqnarray*}
where $x_k^2+y_k^2+z_k^2=1$, and $Z_{\alpha_2}$ denotes a rotation about
the $z$-axis, namely $Z_{\alpha_2}=e^{-i\alpha_2 Z}=(\cos
\alpha_2){\mathbb{I}}  - i (\sin \alpha_2) Z$. 

Let us now check how well the $\ket{z+}$ state is encoded.  
\begin{eqnarray*}
\rho =
\ket{z+}\bra{z+} = \left[\begin{array}{cc}
	1 & 0 \\ 
  0 & 0
\end{array}\right].
\end{eqnarray*}
Note that since $\rho$ is an eigenstate of $Z$, it commutes with
$Z_{\alpha_2}$.  
After some algebraic manipulations we have: 
\begin{eqnarray*}
	I(\rho) &=& \E(\rho)-\half \mathbb{I} 
	\;\;=\;\; \left[\begin{array}{cc}
		\left(w_1+w_2-\half\right) + \left(\sum_{k=3}^n w_k A_k \right) &
	\left(\sum_{k=3}^n w_k B_k \right)^{*} \\  
	  \left( \sum_{k=3}^n w_k B_k \right) & -\left(w_1+w_2-\half\right)
	- \left( \sum_{k=3}^n w_k A_k \right) 
	\end{array}\right], \nonumber
\end{eqnarray*}
where 
\begin{eqnarray*}
	A_k&=&\cos^2 \alpha_k + z_k^2 \sin^2 \alpha_k, \nonumber\\
	B_k&=& (\cos \alpha_k + i z_k \sin \alpha_k) \left(x_k-i y_k\right) i
	\sin \alpha_k. \nonumber 
\end{eqnarray*}
The eigenvalues of $I(\rho)$ are now
\begin{eqnarray*}
	\lambda_{I(\rho)}^2 = \left[\left(w_1+w_2-\half\right) + \left(\sum_{k=3}^n w_k
	A_i\right)\right]^2 + \left|\sum_{k=3}^n w_k B_k
	\right|^2. \label{lambdafor4}  
\end{eqnarray*}
We know that $w_1+w_2 - 1/2 \geq 0$ and  $A_k\geq 0$. Thus we can bound the eigenvalues as
\begin{eqnarray*}
\lambda_{I(\rho)}^2 \geq \left(w_1+w_2-\half\right)^2, \label{lambdamin}
\end{eqnarray*}
with $w_1$ and $w_2$ the two largest weights. 
The equality is achieved if we pick our unitaries with $z_k=0$ and
$\cos\alpha_k=0$, which imply $A_k=B_k=0$.  

This is the same result as in equation (\ref{precision}) and hence
no matter how we pick the unitaries, the encryption cannot be
better than in the Pauli Encryption scheme. \\

\noindent
{\bf Case 3:} $w_1+w_2-1/2 \geq 0$ and $\sum_{k=3}^n w_k \geq 2w_2$. 

Since $w_1+w_2 \geq 1/2$, we conclude that $w_1 \geq 1/4 \geq
\frac{1}{3}(w_2+\sum_{k=3}^n w_k)$ and so, it's possible to consider
the Pauli scheme $P$ that uses the distribution
$\{w_1,\frac{1}{3}(w_2+\sum_{k=3}^n w_k),
\frac{1}{3}(w_2+\sum_{k=3}^n w_k),\frac{1}{3}(w_2+\sum_{k=3}^n w_k)\}$.  
Moreover, the constraint $\sum_{k=3}^n w_k \geq 2w_2$ implies
that $\frac{1}{3}(w_2+\sum_{k=3}^n w_k) \geq w_2$ and hence from the
concavity of the entropy, $P$ uses less entropy than $T$.
From equation \ref{precision}, we know that for the encryption scheme
$P$  
\begin{eqnarray}
\textrm{max}_\rho|\lambda_{I(\rho)}| = \left| w_1+\frac{1}{3}(w_2+\sum_{k=3}^n
w_k)-\half\right|. \label{Pcase3} 
\end{eqnarray}
In what follows, we calculate how well the states $\ket{z+}, \ket{x+}$
and $\ket{y+}$ are encrypted by $T$ and prove that at least one of
them is encoded worse than in the Pauli scheme $P$.

We pick the unitaries of $T$ to be
\begin{eqnarray*}
 {\mathbb{I}}, Z_{\alpha_2}, \left(z_3 Z+ x_3 X+y_3
Y\right)_{\alpha_3},\ldots, 
\left(z_n Z+x_n X+y_n Y\right)_{\alpha_n}.
\end{eqnarray*}
Similarly to Case 2, the $\ket{z+}$ state is encoded no
better than with 
\begin{eqnarray*}
\lambda^2_z \geq \left(w_1 + w_2 + \sum_{k=3}^n w_k (\cos^2\alpha_k+\sin^2\alpha_k\cos^2\beta_k)-\half\right)^2,
\end{eqnarray*}
where we named $z_k=\cos\beta_k$, $x_k=\sin\beta_k\cos\gamma_k$ and $y_k=\sin\beta_k\sin\gamma_k$
Let us now check how well the $\ket{x+}$  state is encoded.   
\begin{eqnarray*}
\rho_x  &=& \half \left[\begin{array}{cc}
	1 & 1 \\ 
  1 & 1
\end{array}\right] \\
Z_{\alpha_2} \rho_x Z^{\dagger}_{\alpha_2} &=& \half \left[\begin{array}{cc}
	1 & e^{-2i\alpha_2} \\ 
  e^{2i\alpha_2} & 1
\end{array}\right] \nonumber \\
U_k \rho_x U_k^{\dagger} &=& \half \left[\begin{array}{cc}
	1+C_k & D_k -i E_k \\ 
  	D_k + i E_k & 1-C_k
\end{array}\right], \nonumber\\
\end{eqnarray*}
where $D_k = ( -1+2\cos^2\alpha_k +2 \sin^2\alpha_k \sin^2\beta_k
\cos^2\gamma_k) $ and $C_k,E_k$ are functions of
$\alpha_k, \beta_k, \gamma_k$ which do not affect the bounds.  
The encoding of $\rho_x$ becomes
\begin{eqnarray*}
I(\rho_x) = \E(\rho_x)- \frac{\ii}{2} &=& w_1 \ii \rho_x \ii + w_2
Z_{\alpha_2} \rho_x Z_{\alpha_2} + \sum_{k\geq 3} w_k U_k \rho_x
U_i^{\dagger} - \frac{\ii}{2} \\ 
&=& \half \left[\begin{array}{cc}
	\sum_{k\geq3} w_k C_k & F-iG  \\ 
  F + i G  & - \sum_{k\geq3} w_k C_k 
\end{array}\right], \nonumber
\end{eqnarray*}
where $F =  w_1-w_2 + 2 w_2 \cos^2\alpha_2 + \sum_{k\geq3} w_k D_k$
and $G = w_2 \sin 2\alpha + \sum_{k\geq3} w_k E_k$.  
We are ready to bound the eigenvalue:
\begin{eqnarray*}
\lambda_x^2 &=& 
\frac{1}{4}\left( \sum_{k\geq3} w_k C_k \right)^2 +\frac{1}{4}G^2
 + \frac{1}{4}F^2 \;\;\; \geq \;\;\; \frac{1}{4}F^2 \nonumber\\ 
& = & \frac{1}{4}\left( w_1-w_2 + 2 w_2 \cos^2\alpha_2  +
\sum_{k\geq3} w_k (-1+2\cos^2\alpha_k + 2\sin^2\alpha_k \sin^2\beta_k
\cos^2\gamma_k) \right)^2 \nonumber\\  
&=& \left(w_1 + w_2 \cos^2\alpha_2 + \sum_{k\geq3} w_k \cos^2\alpha_k
+ \sum_{k=3}^{n} w_k \sin^2\alpha_k\sin^2\beta_k \cos^2 \gamma_k
-\half\right)^2.  
\end{eqnarray*}
Using the same type of computation as above, we encode the $\ket{y+}$
state and obtain a bound for the eigenvalues of
$I(\rho_y)=\E(\rho_y)-\ii/2$:  
\begin{eqnarray*}
\lambda_y^2 &\geq& 
		\left(w_1 + w_2 \cos^2\alpha_2 + \sum_{k\geq3} w_k \cos^2\alpha_k 
			+ \sum_{k=3}^{n} w_k \sin^2\alpha_k \sin^2\beta_k \sin^2 \gamma_k -\half\right)^2. 
\end{eqnarray*}
Summing the three inequalities of the eigenvalues, we obtain that 
\begin{eqnarray*}
|\lambda_x|+|\lambda_y|+|\lambda_z| & \geq & 
\left(3w_1+w_2+\sum_{k=3}^{n} w_k + (2\sum_{k\geq 2} w_k\cos^2\alpha_k) -
\frac{3}{2}\right)\\
& \geq &  \left(3w_1+w_2+\sum_{k=3}^{n} w_k - \frac{3}{2}\right).
\end{eqnarray*}
which implies that at least one of the three $\lambda$ is greater or equal to 
(\ref{Pcase3}). This means the Pauli encryption scheme $P$ is no worse
than $T$, while using less entropy. 

This concludes the proof that
Pauli Encryption schemes are no worse than general encryption
schemes. This also concludes the proof of Theorem 1. 


\section{$N$-qubit independent encryption schemes} 
\label{sectionNqubit}

In this section we consider $n$-qubit encryption schemes
which are composed of independent single-qubit schemes, each using $H$
amount of classical entropy\footnote{ For clarity of exposition, we
assume that for each qubit we use the same amount of classical
entropy. All the results go through in the more general case where for
each qubit $k$ we use entropy $H_k$. }. By independent we mean that the encryption has the form  $R(\rho) = ( R_1 \otimes \ldots \otimes R_n)(\rho)$.

\setcounter{theorem}{1} 
\begin{theorem}
Let $P$ be the single-qubit Pauli encryption scheme, which achieves
the optimal approximation $\ep$ for the given entropy $H$. Then, the optimal $n$-qubit independent encryption scheme $R(\rho)$ is the same for both the 2- and the $\infty$-norm and has the following properties:
\begin{enumerate}
\item $R(\rho) = P^{\otimes n} (\rho)$.  
\item $\normtwo{R(\rho)-\frac{\ii}{2^n}}
\leq \sqrt{n}\frac{\epsilon}{2^{n/2}} +\frac{o(\epsilon\sqrt{n})}{2^{n/2}}.$
\item $\normi{R(\rho)-\frac{\ii}{2^n}}
\leq n\frac{\epsilon}{2^n} + \frac{o(n\epsilon)}{2^n}.$
\end{enumerate}
\end{theorem}

\begin{proof}
We first employ a result by King \cite{King} to show  
that product states are the worst encoded states for independent
encryption schemes. 
King proved that the $p$-norm of a product of unital channels\footnote{A 
quantum channel $\Phi$ is {\em unital} if it preserves unity, 
i.e. $\Phi(\ii)=\ii$. The encryption schemes we consider here are
unital. The $p$-norm of a channel $R$ is the maximum of
$\normp{R(\rho)}$ over all input states $\rho$. Note that the
multiplicativity of the $p$-norms, and hence the additivity of the
capacities of non-unital channels is a main open question \cite{addshor}.}    
is multiplicative, i.e. for $p\geq 1$,
\begin{eqnarray}
	\max_\rho \normp{R(\rho)} = \prod_{i=1}^n \left(\max_{\xi_i} \normp{R_i(\xi_i)} \right).
	\label{kingresult}
\end{eqnarray}
This shows that the norm $\normp{R(\rho)}$ is maximized by a product
state $\rho=\xi_1\otimes\dots\otimes\xi_n$, where $\xi_i$ is the state of the $i$-th qubit. In our encryption
schemes we measure the quality of the approximation by the maximum of
the norm $\normp{R(\rho)-\ii/2^n}$, for $p=2$ and $p=\infty$.   
Let $\lambda_k$ be the eigenvalues of $R(\rho)$; then, the eigenvalues of
$R(\rho)-\ii/2^n$ are $(\lambda_k-1/2^n)$ and we have
\begin{eqnarray}
	\normtwo{R(\rho)-\frac{\ii}{2^n}}^2
		&=& \sum_{k=1}^{2^n} \left(\lambda_k-\frac{1}{2^n}\right)^2	
		= \sum_{k=1}^{2^n} \left( \lambda_k^2-2\frac{\lambda_k}{2^n} + \frac{1}{2^{2n}} \right) =  \normtwo{R(\rho)}^2 - \frac{1}{2^n} , \label{tworelation}\\
	\normi{R(\rho)-\frac{\ii}{2^n}} 
		&=& \max_k \left(\lambda_k-\frac{1}{2^n}\right)	
		= \normi{R(\rho)} - \frac{1}{2^n}. \label{iiirelation}
\end{eqnarray}
It is clear, that the norm of $R(\rho)-\ii/2^n$ is maximized
when the norm of $R(\rho)$ is maximized and
therefore, for any independent encryption scheme the worst encoded
state is a product state. 

Hence, in order to find the optimal independent encryption scheme, one
needs to find the scheme that encrypts product states optimally. 
The encryption of a product state   
$R(\xi_1\otimes\dots\otimes\xi_n)=R_1(\xi_1)\otimes\dots\otimes 
R_n(\xi_n)$ is also a product state and the eigenvalues of 
$R(\xi_1\otimes\dots\otimes\xi_n)$ are
simply products of the eigenvalues of $R_k(\xi_k)$. 
Without loss of generality, let us now encrypt a state
$\xi_1\otimes\rho_{2\dots n}$  
using $R=R_1\otimes R_{2\dots n}$. The eigenvalues of the
single-qubit encryption $R_1(\xi_1)$ can be expressed as  
$\mu_{1,2}=(1\pm\epsilon_1)/2$ and the eigenvalues of $R_{2\dots
n}(\rho_{2\dots n})$ as
$\nu_{k=1,\dots,2^{n-1}}=(1+\delta_k)/2^{n-1}$.  
Hence, the eigenvalues and 2-norm  
of $R(\xi_1\otimes\rho_{2\dots n})-\ii/2^n$ are
\begin{eqnarray}
\label{TensorProductEigs}	\lambda_{i,k} & = & \mu_i \nu_k -\frac{1}{2^n} =  
\frac{\delta_k\pm\epsilon_1(1+\delta_k)}{2^n}, \label{producteigs}\\
	\normtwo{R(\xi_1\otimes\rho_{2\dots n})-\frac{\ii}{2^n}}^2 &
= & \sum_{k=1}^{2^{n-1}} \sum_{i=1}^{2} \lambda_{i,k}^2 =  
		\sum_{k=1}^{2^{n-1}} \frac{\delta_k^2 + \epsilon_1^2
(1+\delta_k)^2}{2^{2n}}. \nonumber
\end{eqnarray}
The last expression is a growing function of $\epsilon_1$ and therefore, the
optimal $n$-qubit encryption scheme has to be optimal (i.e. Pauli) 
on the first qubit, 
giving the smallest possible upper bound on $\epsilon_1$.
After going through this procedure for all qubits, 
we see that the optimal encryption scheme for product states 
in the 2-norm is the Pauli scheme $P^{\otimes n}$. It is 
straightforward to obtain the same statement for the $\infty$-norm 
using \eqref{producteigs}.
This concludes the proof that the optimal $n$-qubit independent
encryption scheme for both the 2- and the operator norm is the Pauli scheme $P^{\otimes n}(\rho)$.  

We now prove tight upper bounds for the quality of the approximation
of the Pauli encryption scheme. 
For the 2-norm, equation \eqref{kingresult} and induction imply
\begin{eqnarray}
	\max_\rho \normtwo{P^{\otimes n} (\rho)} = \left(\max_\xi \normtwo{P(\xi)}\right)^n. \nonumber
\end{eqnarray}
Let us pick $\xi$ to be the worst encoded single-qubit state for $P$. 
The eigenvalues of $P(\xi)$ are $(1\pm\ep)/2$ and therefore:
\begin{eqnarray}
	\normtwo{P^{\otimes n} (\rho)} \leq \max_\rho \normtwo{P^{\otimes n} (\rho)} = \left(  \left(\frac{1+\ep}{2}\right)^2+\left(\frac{1-\ep}{2}\right)^2 \right)^{\frac{n}{2}} 
	= \left(\frac{1+\ep^2}{2}\right)^\frac{n}{2}. \nonumber
\end{eqnarray}
From equation \eqref{tworelation}, we bound the 2-norm of $P^{\otimes n}(\rho)-\ii/2^n$ as
\begin{eqnarray}
	\normtwo{P^{\otimes n} (\rho)-\frac{\ii}{2^n}} =
	\left[ \normtwo{P^{\otimes n}
(\rho)}^2-\frac{1}{2^n}\right]^{\frac{1}{2}}  \leq
		\left[ \left(\frac{1+\ep^2}{2}\right)^n -
\frac{1}{2^n} \right]^\frac{1}{2} 
		= \frac{\ep\sqrt{n}}{2^{n/2}}  + \frac{o(\ep\sqrt{n})}{2^{n/2}} . \nonumber
\end{eqnarray}
For the $\infty$-norm, multiplicativity of norms \eqref{kingresult} implies 
\begin{eqnarray}
		\normi{P^{\otimes n} (\rho)} \leq \max_\rho \normi{P^{\otimes n} (\rho)} 
			= \max_\xi \normi{P(\xi)}^n = \left(\frac{1+\ep}{2}\right)^n, \nonumber
\end{eqnarray}
and therefore equation \eqref{iiirelation} gives us
\begin{eqnarray}
	\normi{P^{\otimes n} (\rho)-\frac{\ii}{2^n}} =
\normi{P^{\otimes n} (\rho)}-\frac{1}{2^n} \leq 
		\left(\frac{1+\ep}{2}\right)^n  - \frac{1}{2^n} 
	 = \frac{n\ep}{2^{n}}  + \frac{o(n\ep)}{2^{n}}. \nonumber
\end{eqnarray}
Note that both bounds are tight and achieved for product states.  
\end{proof}
Since the bounds in Theorem 2 are tight, any good independent encryption scheme 
requires that the approximation parameter for each single qubit is 
$\ep = O(\frac{1}{\sqrt{n}})$ for the 2-norm and $\ep = O(\frac{1}{n})$ 
for the $\infty$-norm. Hence, from equation \eqref{Htwo} we conclude that 
the amount of entropy needed for the encryption of $n$-qubit states is $2n-o(n)$. 


\section{General $N$-qubit Encryption Schemes} \label{sectionNgeneral}

In the previous section, we found the optimal way to independently
compose single-qubit encryption schemes in order to encrypt $n$-qubit
states. However, one can do better with encryption schemes that do not
act independently on each qubit. For example, the encryption scheme 
in \cite{HLSW04} uniformly picks an $n$-qubit unitary from a set of
$O(n2^n)$ random ones and hence it is not an independent
encoding. Note also that it only uses $n+\log n+ O(1)$ bits of
entropy. 

Ambainis and Smith (\cite{AS04}) managed to derandomize the encryption
scheme of \cite{HLSW04} by explicitly describing the set of 
unitaries. In particular, they use a set of $2n$-bit strings, where
each string corresponds to a product of $n$ Pauli matrices (the bits
$\{2j-1,2j\}$ define the Pauli matrix for the $j$-th qubit). They prove
that if the set of strings is a small-bias set of size $O(n2^n)$, then
picking a random unitary from this set gives an $(\epsilon,n+2\log n+
2\log\frac{1}{\epsilon})$ encryption scheme in the 2-norm. 

A $\delta$-biased set is a set of $k$-bit strings such
that for all possible subsets of bits, the probability over the set
that the parity of the subset is 0, is
$[\frac{1}{2}-\delta,\frac{1}{2}+\delta]$.    
Naor and Naor \cite{NN93} gave the first such construction with size
polynomial in $k$ and $1/\delta$. Alon, Goldreich, Hastad and Peralta 
\cite{AGHP92} showed a lower bound on the size of a $\delta$-biased set
\[ N(k,\delta)\geq  \Omega\left(\min\left\{\frac{k}{\delta^2 \log(1/\delta)},
2^k\right\}\right)\] 
Since we are interested in encryption schemes which use less than 2n
bits of entropy, we only consider $\delta$-biased sets of size
$o(2^{2n})$ and hence $\delta = \omega(\frac{1}{2^n})$. 
Ambainis and Smith showed the following:
\begin{quote}
There exists a function $\delta(n) = \omega(\frac{1}{2^n})$ such that
any $O(\delta(n))$-biased set gives rise to a good encryption scheme
in the 2-norm and moreover it has size $N=o(2^{2n})$. 
\end{quote}
In fact, their result holds for any $\delta(n)=
\frac{1}{\alpha(n)2^{n/2}}$, where $\alpha(n)$ is 
any slowly growing function of $n$ (e.g. $\log n$).
Note, that there are explicit constructions of such small-bias sets
of size $N = \textrm{poly}(\alpha(n),n)2^n$.    
However, it was an open question whether the same holds for the case
of $\infty$-norm. 
Here, we resolve this question by providing a
counterexample. We show that
\begin{quote}
For any $\delta(n) = \omega(\frac{1}{2^n})$
there exists a $O(\delta(n))$-biased set of size $N=o(2^{2n})$ which is not
good in the $\infty$-norm.
\end{quote}
Let us, first, compute the norm $\normi{R(\rho)-\frac{\ii}{2^n}}$, where
$R$ is a Pauli encryption scheme and
$\rho = \ket{z+}^{\otimes n}\bra{z+}^{\otimes n}$. The
density matrix of this state in the $z$-basis is
\begin{eqnarray*}
\rho=\left(\frac{\ii+Z}{2}\right)^{\otimes n}=\left(
\begin{array}{ccc}
	1 & 0 & \dots \\
	0 & 0 & \dots \\
	\vdots & \vdots & \ddots
\end{array}\right).
\end{eqnarray*}
The unitaries in the encryption scheme can be written as $U^k
= U^k_1 \otimes \dots\otimes U^k_n$ with $U^k_i\in\{\ii,X,Y,Z\}$.  
For each unitary $U^k$ we define a string $\chi^k\in\{0,1\}^n$  with
$\chi^k_i=0$ if $U^k_i\in\{\ii,Z\}$,  and $\chi^k_i=1$ if
$U^k_i\in\{X,Y\}$. Note that 
$XZX=YZY=-Z$ and $ZZZ=\ii Z\,\ii=Z$, and hence
\begin{eqnarray*}
	R(\rho) = \frac{1}{N} \sum_{k=1}^{N} 
\left(\frac{\ii+(-1)^{\chi^k_1} Z}{2}\right)\otimes \dots
\otimes \left(\frac{\ii+(-1)^{\chi^k_n} Z}{2}\right).
\end{eqnarray*}
The density matrix of the encrypted state is again diagonal in the
$z$-basis and therefore, its eigenvalues are simply its diagonal
elements. The size of each eigenvalue $\lambda_\chi$ is exactly
the number of unitaries $U^k$ with the same corresponding string
$\chi$ divided by $N$. Thus,
\begin{eqnarray*}
\normi{R(\rho)-\frac{\ii}{2^n}}=\max_\chi \left|\lambda_\chi
-\frac{1}{2^n}\right| = 
\max_\chi \left| \frac{1}{N}(\# \textit{of unitaries with the same
}\chi)-\frac{1}{2^n}\right|. 
\end{eqnarray*}

It is easy to see that starting from any small-bias set we
can create a set which is asymptotically as good as the initial one
and it has the extra property that it contains at least a $\delta(n)$
fraction of the unitaries with $\chi=0$. We start with a
$O(\delta(n))$-biased set of size $N$ and add $\delta(n) N$
unitaries with $\chi=0$ to the initial set. The new set has size $N' =
O(N)$ and bias $O(\delta(n))$, and therefore, it is asymptotically 
as good as the original set. Hence
\[ \normi{R(\rho)-\frac{\ii}{2^n}} \;=\; \max_\chi \left| \frac{1}{N'}(\#
\textit{of unitaries with the same 
}\chi)-\frac{1}{2^n}\right| \;\geq\; O(\delta(n))-\frac{1}{2^n} \;=\;
\omega\left(\frac{1}{2^n}\right),\]   
which means that the encryption scheme $R$ is not good in the
$\infty$-norm.
In other words, we show that although a $\delta$-biased set encryption
scheme is always good for the 2-norm, this is not the case for the
$\infty$-norm. However, it is still conceivable that one might be able
to use $\delta$-biased sets with some extra properties in order to
achieve good encryption for the $\infty$-norm.


\section*{Acknowledgements}
We would like to thank Eddie Farhi and Debbie Leung for stimulating discussions, 
and the journal referees for pointing out references \cite{Nielsen1} and \cite{Ruskai} and 
for the simplified proofs in Section II.1 and Appendix A.2.
IK  acknowledges professor Peter w. Shor who has
supported his research from his
appointment as the holder of the H. A. Morss and H. A.
Morss, Jr. Professorship and from NSF CCF-0431787.
DN gratefully acknowledges support from the National Security Agency (NSA) 
and Advanced Research and Development Activity (ARDA) 
under Army Research Office (ARO) contract W911NF-04-1-0216.


\appendix

\section{Appendix}
\subsection{Minimization of the entropy $H(\epsilon,z)$}
Here, we provide the details for the minimization of the function $H(\epsilon,z)$ with respect to $z$ that concludes the proof of the optimal trade-off between approximation and entropy. 
Recall that
\begin{eqnarray*}
H(\ep,z)=-\left(\half+\ep-z\right)\log\left(\half+\ep-z\right)
-2z \log z -\left(\half-\ep-z\right)\log\left(\half-\ep-z\right).\nonumber
\end{eqnarray*}
and the constraints are 
\begin{eqnarray}
\frac{1}{4}+\frac{\ep}{2} &\geq& z \geq \frac{1}{4}-\frac{\ep}{2},\\ 
\half-\ep &\geq& z.
\end{eqnarray}
The entropy as a function of $z$ is again a concave function
and hence, in order to find the minimum we investigate the endpoints
of the allowed interval for $z$. There are two cases: \\
\noindent
{\bf Case 1:} for $\ep \leq 1/6$, the constraint (\ref{zconstraint1}) is
tighter. The left endpoint is
$\{\frac{1}{4}+\frac{\ep}{2},\frac{1}{4}+\frac{\ep}{2}, 
\frac{1}{4}+\frac{\ep}{2}, \frac{1}{4}-\frac{3\ep}{2}\}$, giving 
\begin{eqnarray*}
\quad H_1(\ep)
&=&-3\left(\frac{1}{4}+\frac{\ep}{2}\right)\log\left(\frac{1}{4}+\frac{\ep}{2}\right)
-\left(\frac{1}{4}-\frac{3\ep}{2}\right)\log\left(\frac{1}{4}-\frac{3\ep}{2}\right) \\
&=& 2-\frac{6}{\ln 2} \ep^2  +\textsl{O}(\ep^3) \label{Htwo} 
\end{eqnarray*}
The right endpoint is $\{\frac{1}{4}+ \frac{3\ep}{2}, \frac{1}{4}-
\frac{\ep}{2},\frac{1}{4}- \frac{\ep}{2},\frac{1}{4}- \frac{\ep}{2}
\}$, giving 
\begin{eqnarray*}
\quad H_2(\ep) &=&
-\left(\frac{1}{4}+\frac{3\ep}{2}\right)\log\left(\frac{1}{4}+\frac{3\ep}{2}\right)
-3\left(\frac{1}{4}-\frac{\ep}{2}\right)\log\left(\frac{1}{4}-\frac{\ep}{2}\right)\\
\end{eqnarray*}
At $\ep=0$, $H_1=H_2=2$. At $\ep=1/6$, $H_1 \leq H_2$. The derivative of $H_2-H_1$ is always negative,
\begin{eqnarray*}
\frac{\textrm{d} (H_2-H_1)}{\textrm{d} \ep}= 
\frac{3}{2} \log \frac{ \left(1-2\ep\right)\left(1-6\ep\right)}{\left(1+2\ep\right)\left(1+6\ep\right)} \leq 0,
\end{eqnarray*}
so we conclude that $H_1$ is the best choice for $\ep\leq 1/6$. At $\ep=1/6$, $H_1$ achieves the value of $\log_2 3$, which means only three 
equally weighed unitaries are used.\\

\noindent
{\bf Case 2:} for $\ep\geq 1/6$, the constraint (\ref{zconstraint2}) is tighter,
changing the left endpoint of $z$ to 
$z=1/2-\ep$. This sets $y=0$, which is the regime of using only three
unitaries, i.e. the distribution is $\{2\ep,\half-\ep,\half -\ep,0\}$
and the entropy  
\[
H_3(\ep) = -2\ep\log 2\ep -
2\left(\frac{1}{2}-\ep\right)\log\left(\frac{1}{2}-\ep\right).\]
The second derivative of $H_3-H_2$ is always negative,
\begin{eqnarray*}
\frac{\textrm{d}^2}{\textrm{d} \ep^2} (H_3-H_2)= 
-2 \left[\ep ( 1 - 2\ep) ( 1 + 6\ep) \ln 2 \right]^{-1},
\end{eqnarray*}
so the function $H_3 - H_2$ is concave. That allows for only two
points where $H_3=H_2$. One of them is at $\ep=1/2$, the other is
found numerically to be $\ep_0 \approx 0.287$ with $H_0\approx
1.41$. We conclude that for $1/6 \leq \ep \leq \ep_0$, the choice of
$H_3$ is optimal, whereas for $\ep_0 \leq \ep \leq 1/2$, the best
choice is $H_2$. 


\subsection{Another proof of optimality of Pauli Encryption Schemes}
It is known \cite{Ruskai} that every unital channel $\E$ with weights $\{w_k\}$
and unitaries $U_k$ is equivalent to a Pauli channel with 
some other weights $\{x_m\}$, that is
\begin{eqnarray}
	\E(\rho) &=& \sum_k w_k U_k \rho U_k^\dagger, \\
  \E(\rho) &=& x_1 \rho + x_2 X\rho X + x_3 Y\rho Y + x_4 Z\rho Z.
\end{eqnarray}
This channel is an $\epsilon$-randomizing map. We will prove
that the Pauli realization of it has smaller entropy. 

Suppose we act with this channel on one half of the Bell state 
$\kets{\psi^+}=\frac{1}{\sqrt{2}}\left(\ket{00}+\ket{11}\right)$.
The first definition of $\E$ will give us
\begin{eqnarray}
	\rho' = \E(\rho) = \sum_k w_k \left(U_k\otimes \ii\right) 
			\kets{\psi^+}\bras{\psi^+} \left(U_k\otimes \ii\right)^\dagger 
			= \sum_k w_k \ket{\psi_k}\bra{\psi_k},
\end{eqnarray}
where $\ket{\psi_k}=\left(U_k\otimes\ii\right) \kets{\psi^+}$ are pure states.
On the other hand, the second realization of $\E$ (with Pauli operations)
acting on one half of the state $\kets{\psi^+}$ will transform 
it into a state with density matrix diagonal in the Bell basis,
$\rho' = \textrm{diag}(x_1,x_2,x_3,x_4)$.

In \cite{Nielsen1}, Nielsen showed that when a density matrix 
can be expressed as $\rho'=\sum_k w_k \ket{\varphi_k}\bra{\varphi_k}$,
where $\ket{\varphi_k}$ are normalized states, the (ordered) vector of
probabilities $w_k$ is majorized by the vector of eigenvalues of 
$\rho'$, that is $(w_k) \prec \lambda(\rho')$.

In our case, the vector of eigenvalues of $\rho'$ is $(x_m)$,
and the majorization $(w_k) \prec (x_m)$ 
means $\sum_{m=1}^{n} w_m \leq \sum_{m=1}^{n} x_m$ for any $n\geq 1$.
Note that if the length of $(w_k)$ is greater than four, we pad
the vector $(x_m)$ by zero entries to make the lengths of the vectors equal.

The entropy function is concave. Because the
vector of weights for the Pauli realization $(x_k)$
majorizes the vector of weights for the original realization $(w_k)$, 
the Pauli realization of the channel has smaller entropy, 
$S(\{x_k\})\leq S(\{w_k\})$.
This means the Pauli channel is the optimal (entropy-wise)
realization of any unital channel.

\end{document}